# Forensic Considerations for the High Efficiency Image File Format (HEIF)

Sean McKeown
*School of Computing*
*Edinburgh Napier University*
Edinburgh, Scotland
s.mckeown@napier.ac.uk

Gordon Russell
*School of Computing*
*Edinburgh Napier University*
Edinburgh, Scotland
g.russell@napier.ac.uk

*Abstract*—The High Efficiency File Format (HEIF) was adopted by Apple in 2017 as their favoured means of capturing images from their camera application, with Android devices such as the Galaxy S10 providing support more recently. The format is positioned to replace JPEG as the de facto image compression file type, touting many modern features and better compression ratios over the aging standard. However, while millions of devices across the world are already able to produce HEIF files, digital forensics research has not given the format much attention. As HEIF is a complex container format, much different from traditional still picture formats, this leaves forensics practitioners exposed to risks of potentially mishandling evidence. This paper describes the forensically relevant features of the HEIF format, including those which could be used to hide data, or cause issues in an investigation, while also providing commentary on the state of software support for the format. Finally, suggestions for current best-practice are provided, before discussing the requirements of a forensically robust HEIF analysis tool.

## I. INTRODUCTION

Modern technology is ever changing, often developing at a radical pace. This frequently creates challenges for law enforcement agencies, necessitating the development of new techniques to handle digital evidence as part of an ongoing technological arms race. In contrast, one area which has not seen much movement is the most frequently used mechanism for storing image files. The JPEG standard [1] has been the de facto image compression scheme for decades, with all but a monopoly on mobile, desktop and server devices alike. However, the JPEG standard was published in 1992, and as such has become rather antiquated, lacking support for modern features, such as high colour depth, high dynamic range (HDR), and transparency, while producing visual artefacts and effecting sub-optimal compression. While there have been several attempts to replace JPEG with technically superior alternatives, such as Better Portable Graphics (BPG) [2], WebP [3], and several extensions and reworks of JPEG[1], none have succeeded. The user base is so entrenched in JPEG, due to its universal software support, that it is very difficult for a new standard to gain enough of a critical user mass to trigger large scale migration. However, this may change with the relatively recent adoption of the High Efficiency Image File Format (HEIF) by Apple in 2017, beginning with the iPhone 7, 6th generation iPad, and Mac computers running OS X High Sierra (10.13) and above. While claims of higher encoding efficiency have done little to sway users in the past, Apple's hardware and software support across millions of devices potentially allows the HEIF format to eventually replace JPEG in mainstream use. Contemporary Apple devices save images from the camera directly in the HEIF format by default. This sudden switch took many users by surprise as the lack of software support at the time resulted in problems viewing and sharing images outside of the Apple device ecosystem. Many users find themselves exporting images in the JPEG format for compatibility reasons, even years after the initial launch of the iPhone 7 [4].

Unfortunately this lack of widespread software support is also an issue for forensic investigators, as many forensics tools have also been slow to implement decoders. Two reasons for this is that HEIF is a complex container format which allows multiple images, or burst shots, to be stored in many potential configurations, while also lacking mature publicly available implementations or documentation. This presents a challenge for the forensic investigator, as iOS devices have an approximate global market share of 25% at the time of writing [5], meaning that 1 in 4 investigations involving a mobile phone can potentially encounter HEIF images. Other manufacturers may also follow Apple's lead, with Android supporting HEIF from Android 9 (Pie) and Samsung presenting users with the option to save photos as HEIF files on the Galaxy S10 [6].

This paper discusses the forensically relevant features of the HEIF format and surveys existing tool support. Several features of the format can be used to hide data, such that it is critical that examiners are aware of the capabilities of this container and how best to approach its analysis. The latter parts of the paper will therefore outline a recommended forensics process given the state of existing software support before providing a set of requirements which should be implemented by a forensic HEIF viewer.

## II. THE HIGH EFFICIENCY FILE FORMAT

The HEIF format, defined in ISO/IEC23008-12 [7], is built upon the ISO Base Media File Format (ISOBMFF) [8], which is an extensible format designed for video and audio content as part of the MPEG–4 part 12 specification. As it is a media container format, HEIF is not only able to store multiple images, but also image sequences, such as burst shots, or animations. Still images are typically compressed using intra-coding techniques from the High Efficiency Video Coding (HEVC, aka H.265) standard, which uses various forms of prediction to reduce redundancy, not unlike existing standards, such as PNG. HEVC encoding has been shown to reduce the number of bits required by approximately 50% over JPEG, and by approximately 25% over more advanced schemes, such as JPEG 2000 [9], [10]. The format also has support for more traditional image codecs, such as JPEG, however this would waive any advantages in file size. When encoding image sequences, temporal, or inter-frame, coding is used, which effectively leverages the video encoding features of HEVC to predict the differences between adjacent frames, resulting in heavy bit rate reductions of similar sequences encoding using intra-coding [11].

---
[1] https://jpeg.org/index.html

HEIF files are composed of basic data structures called boxes, which have a named mnemonic of four ASCII characters, followed by the size of the box in bytes, and a payload. While this is generally similar to chunks in the PNG format [12], HEIF boxes can be nested, creating a hierarchy of boxes and relationships between boxes. As the format was extended from ISOBMFF, which is used in files such as MP4, HEIF more closely resembles a video file format than a traditional image file format. HEIF is therefore a complex container format containing many box types, such that a full analysis of the box hierarchy is outwith the scope of this work. Existing documentation for the format [11], [13] and the ISO standard [7] contains more detail. However, for context, a brief description of a typical HEIF file is provided below.

The first box which appears in the file is `ftyp`, which contains general encoding metadata for the file, depicted as 'brands'. The brand places restrictions on the coding format, with 'mif1' specifying a still image where any codec can be used, with an equivalent 'msf1' for image sequences, both of which correspond to the *.heif* file extension. Brands corresponding to the *.heic* file extension all make use of HEVC encoding, with the specification of a specific encoding profile. These are 'heic'/'heix' for still images, and 'hevc'/'hevx' for image sequences. It should be noted that there is no requirement in the ISOBMFF or HEIF specifications for a magic number (file signature) at the very beginning of the file, though this may be optionally specified in some conforming ISOBMFF files. HEIF files begin with a header for the `ftyp` box, specifying the box length, such that the best way to determine if a file is HEIF is to inspect this box and its brands.

A `meta` box then contains the remaining metadata, beginning with the handler specifier, `hdlr`, which is always 'pict' for HEIF, but may differ in other file formats. The location of each image, or image sequence, with the file is then mapped using the `iloc` box, which acts as an index to specify item offsets and lengths. All images in the file are then described in the item information, `iinf`, box, which contains a separate item entry, `infe`, for each image. An `iref` box is then used to form relationships between images, such as linking a thumbnail (`thmb`) to a full sized image, or metadata stored in the EXIF, XMP or MPEG-7 formats. When multiple images are present in the file, a single image can be tagged as the primary item (`pitm`), meaning that it should be the default image to display in a viewer.

HEIF images can also be assigned properties, contained within an item properties box (`iprp`). These properties can be descriptive, such as the colour profile (`colr`) or aspect ratio (`pixi`), but in stark contrast to traditional image formats, they can also be transformative properties. This is achieved by using a *derived* image, which is created by manipulating one or more *master* images, which serve as a base image. This means that the underlying original is not actually modified, allowing for lossless editing, presenting the manipulated version to the user. These transformations are not arbitrary, however, and must be supported by the file reader. They are: rotate (`irot`), crop (`clap`), and mirror (`imir`). It is also possible to render multiple images together, presenting it as a single image, such as in a grid (`grid`) or overlay (`iovl`). Alternatively, non-displayable auxiliary images (`auxc`) can be included to be used as depth maps or alphas masks for transparency and blending multiple images together (also `iovl`). In each case, the derived image has its own `infe` entry, which specifies the type of derived image, such as grid. These derived images are linked with their master images and auxilliary images using the appropriate references in the `iref` box, which is also the case for auxilliary images. Derived images can also serve as the primary item (`pitm`) to serve as the main display image.

Image sequence data is primarily contained in a movie box, `moov`, which in other file types is used to contain video content. The movie box is not nested within the `meta` box, and contains one or more track (`trak`) boxes, each of which contains an image sequence. The `trak` box also has a `hdlr` box, which again can only be set to 'pict' in HEIF, distinguishing the content from video files in other formats. There are many other elements in the track relating to media metadata, and data handling, which will not be detailed here.

One further feature of note is that HEIF has support for image tiling, which is part of the HEVC specification [14]. Tiles allow for larger images to be pieced together from smaller chunks, though not quite in the same way as the grid functionality. This allows for each image segment to be decoded in parallel, or for small portions of very high-resolution images to be decoded in isolation, without decoding the full image. Tiles enable faster processing and reduced memory loads in certain circumstances.

At the very end of the file is the compressed data region, contained in either an image/item data (`idat`) box or media data (`mdat`) box for still image content, while sequences can only be stored in an `mdat` box.

*A. Implementations*

Unlike the JPEG working group, the team behind the ISO specification for HEIF have not provided an implementation.

```
....ftypmifl....        ....ftypheic....
miflheic...±meta        miflheic...Wmeta
........!hdlr....       ........"hdlr....
....pict........        ....pict........
........pitm...         ........$dinf..
..ô...tiloc....D        ..dref..........
@...ę.......É...        ..url........pi
.....^).í.......        tm.....%...Aiinf
É....^1...Õ.đ...        .....'....infe..
....É....y...sł.        ......hvc1.....i
ó.......É....ěą.        nfe........hvc1.
.‹Ñ.ô...........        ....infe........
..........¬iinf.        hvc1.....infe...
..........infe...       .....hvc1.....in
..ę..hvc1HEVC.Im        fe........hvc1..
age.....infe....        ....infe........h
.í..hvc1HEVC.Ima        vc1.....infe....
ge.....infe.....        ....hvc1.....inf
đ..hvc1HEVC.Imag        e........hvc1...
e.....infe.....ó        ..infe........hv
..hvc1HEVC.Image        c1.....infe.....
...."infe.....ô.        ...hvc1.....infe
.iovlDerived.ima        ........hvc1....
ge.....iref.....        .infe........hvc
...dimg.ô...ę.í.        1.....infe......
                        ..hvc1.....infe.
                        .......hvc1.....
    Nokia overlay           iPhone tiled
```

Fig. 1. **ANSI header view of the overlay_1000x680.heic from the Nokia dataset (left - 4 master images, 1 derived overlay), and a sample from the iPhone XS Max (right), which uses many small image tiles.**

At the time of writing, the most general purpose implementations appear to be libheif [15] and the Nokia implementation [16], the foundations of which are documented by Heikkilla [13]. Both of these implementations are written in C++ and available on Github.

Apple announced the details of their implementation, and adoption of the HEIF format, at their WWDC conference in 2017. Apple use the file extensions .heic/.heics for HEVC encoded still images and image sequences, respectively, with .acvi/.avcs being used for images encoded using the older H264 codec. Files using neither of these have the .heif/heifs extension, which matches the behaviour of the standard [17]. Apple's box layout for still images is `ftyp`, followed by a `meta` box, and lastly an `mdat` box. The `mdat` always contains EXIF metadata, followed by a 320×420 pixel thumbnail, and the compressed image data. Apple makes use of 512×512 tiles for all images, resulting in many `infe` boxes being used for each tile, for a single image. Tiles are then bound together in a single image using a derived `grid` image. Apple also makes use of depth maps (`auxc`), which are accompanied by XMP metadata.

The structure for image sequences is very similar, except that the `mdat` box is followed by a `moov` box containing `trak` items for video, audio and images, with the latter potentially specifying a sequence of thumbnails [17]. There is also the possibility of providing timing information to determine how long each image in a sequence is displayed. However, Apple does not appear to make use of the HEIF format for its live photos feature, which appear to default to a sequence of JPEGs accompanied by a Quicktime movie file. Users can, however, enable HEVC/HEIF such that live photos are saved in the HEIF format, but still have an accompanying Quicktime file. Apple's desktop devices have supported HEIF since High Sierra (10.13), and it has been used to implement the dynamic wallpaper feature of macOS Mojave (10.14) [18].

### III. EXISTING WORK ON HEIF FORENSICS

Despite being adopted by Apple in 2017, it doesn't appear to have attracted a great deal of attention from the digital forensics research community, with the authors failing to identify any existing peer-reviewed work on the format. However, there has been coverage on technical blog posts. In 2017, Leong [19] provided an early first look from a forensics perspective, noting that the format is not widely supported by software in general. Still images were found to be supported by Dropbox's preview system, while HEVC .mov files can be processed by *FFmpeg* to extract frames or transcode files, or played using VLC or Infranview. An analysis of an image from an iPhone 8 is provided, which matches the usage of the format described by Apple [17]. Apple's HEIF and MOV files are noted to store data in big-endian format. Prasannan [20], of CCL forensics, noted that early support for the format was provided in *Exiftool*, Mobil Edit Forensic Express 4.2, and a plugin for Xnview. A year later, Faulkner [21], from Paraben forensics, described poor support by Apple's own photo viewer applications, and the lack of ability to export files in certain configuration. The author again notes the use of cloud services to preview images, except in cases where this is inappropriate, or would distribute illegal media.

More recently, in 2019, Krawetz [22] noted that there is still a lack of publicly available information on the format which is palatable to non-specialists. Similarly, tool support was still lacking, with no browsers supporting HEIF, and poor support by forensics tools. Given the lack of tool support, and motivation to avoid using cloud provider previews, the suggested approach to HEIF image forensics is often to export them to the JPEG format [20], [21].

#### A. Contemporary Preview Support

Fortunately the situation for previewing HEIF images has improved. Research carried out for this paper, which will be expanded on in Section IV, has confirmed that support for viewing HEIF images on macOS is greatly improved and consistent, while Windows has provided support for the format since version 1809, however there is some work required from the user, as the Windows Store extensions 'HEIF Image Extensions'[2] and either 'HEVC Video Extensions from Device

---

[2]https://www.microsoft.com/en-gb/p/heif-image-extensions/9pmmsr1cgpwg?activetab=pivot:overviewtab

*Manufacturer"*'[3] (free) or *'HEVC Video Extensions'*[4] ($0.99) are required. After installing both extensions, the Windows Photos viewer is able to preview them, and thumbnails are generated in Explorer. These extensions were not installed by default with a standard installation of Windows 10 Professional version 1909. However, Linux support does not appear to have caught up, as HEIF is not supported by default on Ubuntu 19.10, and is not provided by the third party media extensions which is an optional package at installation time. However, this may be due to licensing concerns, rather than a lack of technical support.

Despite improved Operating System support for the HEIF format, simple previewing is not sufficient for a forensic investigation, as the complex container format requires a more detailed analysis. The following section discusses some of the forensically relevant features of the format, before analysing tool support for exploring these properties with existing software.

## IV. Forensic Considerations and Software Support

This section discusses the forensically relevant features and software support for the HEIF format, derived from an analysis of the ISO document [7], sample images from the Nokia reference set [16], and sample images from an iPhone XS Max. The discussion will highlight difficulties in forensically analysing the format and the potential for evidence to be hidden within the container format. The Nokia images conform to the ISO specification document, meaning that they are good candidates for experimentation, while the iPhone images represent Apple's implementation which is already in circulation worldwide. Several images were also modified using the Nokia API [16], the code for which is available on Github [23].

### A. General Software Support Overview

Software support for the HEIF format is patchy, with many utilities still not supporting any aspect of the format. For example, the popular image editing software Adobe Photoshop and Lightroom only support the HEIF format on macOS, with no support on Windows even on the latest CC 2020 release. Similarly, a variety of forensics tools (X-Ways Forensics 19.8, Encase 8.07, FTK Imager 4.2.1.4, Autopsy 4.13[5]), were tested and demonstrated to have no support at all for viewing HIEF images. Table I summarises the HEIF support for a selection of HEIF supporting software tested in this work.

In testing, it was noted that the macOS Preview appears to have increased support between Mojave (10.14) and Catalina (10.15), primarily in its ability to view derived images and image sequences. Windows 10 only has support with additional packages installed, with the behaviour depending on which version of the HEVC video extension (discussed above in Section III-A) is installed, with the paid version offering slightly better support than the free version. The Windows 10 build does not appear to have an impact on this behaviour, and both Explorer and the Photo application show consistent support. On non-Mac devices, the standout tool for viewing HEIF files appears to be GIMP, which showed excellent support for both still images and image sequences, while FFmpeg appears to be the most complete Windows solution for handling image sequences.

What follows is a more in-depth discussion of the various HEIF features, with more detailed reporting of software support for handling these features.

### B. Multiple Images

The most salient property of HEIF which differentiates it from popular file formats such as JPEG, BMP, and PNG, is that it is a container which can hold multiple images, or sequences, each with their own associated metadata. This is quite different to the multi-frame animations possible in GIF or APNG[6], as items in HEIF need not be part of the same image stream, and can represent an entire gallery of completely independent images. As such, HEIF should not be approached as a single, static image. Experiments with several software packages which support HEIF showed that, in most cases, only the cover image, which has been marked by the primary item reference (pitm) is typically displayed. This was the case for Windows Explorer, Windows Photo Viewer, Dropbox's Web Preview, Photoshop (Mac) and HEIF to JPEG conversion tools. This means that if a HEIF file contained multiple images and was viewed using one of these tools, or converted to JPEG, only a single image would be displayed, meaning that illegal media could easily be concealed. However, the GIMP[7] image editor and macOS Preview utility, on both Mojave (10.14) and Catalina (10.15), have support for viewing multiple images contained within a HEIF file. This problem extends to viewing the metadata of the images, as both Exiftool and CopyTrans [8] only displayed metadata for the primary image. Figure 2 depicts a screenshot of a multi-image HEIF file (with a derived grid) rendering in the macOS Preview utility.

### C. Embedded Thumbnails

One or more images, or indeed image sequences, within the HEIF container may contain a thumbnail. When the primary image contains a thumbnail, there is the option of displaying the thumbnail as a preview (for example in Windows Explorer), rather than generating one from the full sized image. In our testing the only software which was found to support viewing a thumbnail in a HEIF file was the CopyTrans application, which also uses it as the preview in its Explorer equivalent folder preview. No other software was able to display embedded thumbnails, which is a clear gap in functionality. As the content of a thumbnail need not match the content of the master image it is attached to, this offers an easy opportunity to conceal contraband material.

---

[3]https://www.microsoft.com/en-us/p/hevc-video-extensions-from-device-manufacturer/9n4wgh0z6vhq

[4]https://www.microsoft.com/en-us/p/hevc-video-extensions/9nmzlz57r3t7

[5]While not tested, there does not appear to be any evidence that Forensic Toolkit (FTK) supports HEIF.

[6]Animated PNG

[7]GIMP presents an image selection window, showing all visible items in the container, before opening one specific image for editing.

[8]https://www.copytrans.net/copytransheic/

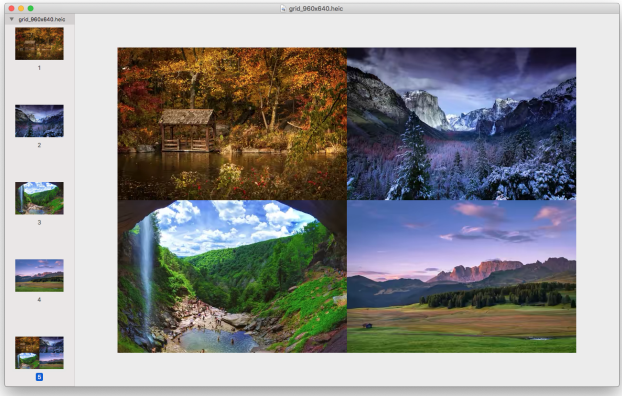

Fig. 2. **The macOS Preview (Mojave 10.14) utility depicting all four component images and the derived grid image for Nokia's grid_960x640.heic file.**

### D. Lossless Modification and Auxiliary Images

One of the key features of HEIF is lossless modification of the underlying image, which can take the form of cropping, rotating, mirroring, or applying depth or transparency masking via auxiliary images. In all cases a derived image is generated, which may optionally be set as a primary image. Software support for the grid view and cropping appears reasonable, while few tools support rotation and mirroring (see Table I). In no instance was an auxiliary image viewable by any of the tools. It is important that these features are understood, as while individual images may be innocuous, their combination may be used to reconstruct illegal media. Take for instance a depth map, which contains an image of greyscale intensity values, which could be used to conceal a greyscale version of illegal media. Similarly, the tiling functionality may be abused to display one image, but allow the reconstruction of another. Another example may be overlaying images using alpha masks to reconstruct an image, much in the same way that separate RGB (red-green-blue) colour channels are combined to make a colour image.

### E. Image Sequences

The tested software, largely aimed at rendering still images, had poor support for HEIF image sequences, either in the burst shot format, or those intended to be consumed as animated images in the same manner as GIF animations. GIMP was the only still image tool which appeared to have reasonable support for image sequences, displaying cover images (first frames) for both types of image sequence, but no further frames for GIF style animations. However, for burst shots, GIMP was able to display a frame for each `infe` item contained in the `meta` box, though these appear to be preview snapshots which may not be present in all files containing image sequences, as image sequence data itself is stored in the `moov` box. In the case of the burst shot images in the Nokia test set, this corresponded to four `infe` box images, resulting in a reasonable preview. Much better support for image sequences is provided by FFmpeg, which effectively has full support for the HEVC inter-coded contents, but does not decode still images as it requires a `moov` box to be present. Individual frames can be extracted from the sequence as JPEG or PNG using the command `ffmpeg -i <filename> <outname>%03d.<extension>` Other video players, such as MPC are able to play image sequences as if they were a video file. Attempts to carve HEVC sequences from the format resulted in limited success with Scalpel and Foremost, two popular data carving tools, as the sequences were unplayable in video players.

It should be noted that there is the potential to hide images in image sequences by altering their timing properties, however this attack should only be possible when the player respects, or implements, manually specified timings.

### F. Hidden Image Flags

Each image in HEIF can be assigned a role, such as cover (primary), thumbnail, auxiliary, or derived, as already discussed. However, an image may also be flagged as hidden, such that it is not intended for display purposes. This could be used legitimately because the image is part of the foundation of a derived image and is not designed for user consumption, however it could also be trivially used to hide images in a HEIF player. Each HEIF box has a standard format, where the first four bytes correspond to the ASCII name of the box, followed by one byte for version number (which was 0x02 for all test images), and three more bytes to set flags, a format which is unmodified from the ISOBMFF standard [8]. The hidden flag can be set in an individual `infe` box by manipulating a single bit in the flags field, as highlighted in Figure 3. This flag appears to be respected by all of the applications tested in this work (Table I), meaning that without analysing the file in a hex editor, or via other means, hidden content would easily be missed in an investigation. It should be noted that hiding the cover image is not supported by the standard, and was ignored in these cases. Setting auxiliary images to hidden caused some tools to report file corruption (macOS Preview and Photoshop on macOS). Individual `infe` items can also be hidden, which is respected by image viewers, such as GIMP and macOS Preview, but is ignored by video based utilities, such as Media Player Classic Home Cinema (MPC) and FFmpeg. Figure 4 depicts the GIMP image selection window for a burst shot image sequence, illustrating what happens when an `infe` box is set to hidden. Images taken with the iPhone, which makes use of a grid, have the hidden bit set for all tile images, such that only the main derived image is displayed.

It is also possible for image sequence tracks (`trak`) to be set inactive using flags in the base ISOBMFF standard [8]. In this case the flags are set in the track header (`tkhd`) box, with a value of 0x00 indicating disabled, 0x01 indicating enabled, another value for when the track is used in presentation (0x02), or in preview only (0x04). This flag did not seem to affect the output of FFmpeg or MPC for single track HEIF files in our tests.

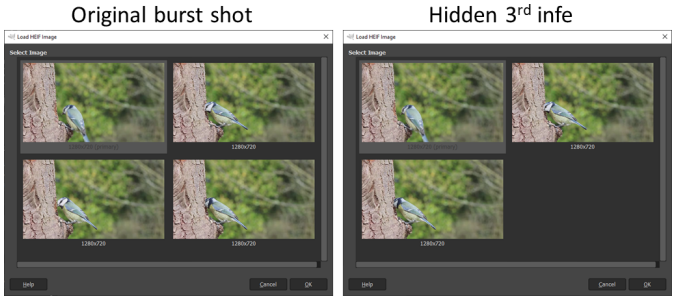

Fig. 3. **Hex editor comparison of an image information entry (infe) box where the item is set to visible (top) and hidden (bottom). Entries differ by a single bit.**

Fig. 4. **The GIMP image selection view for Nokia's bird_burst.heic (left) and the same file modified to hide the third `infe` item (right).**

### G. External Content References

The `meta` box can contain a data information `dinf` box, which itself contains a data reference `dref` box, which specifies the sources in URL or URN (Uniform Resource Name) format [8]. This field can be used to specify data for the file which is not contained within the file itself, as an external reference. This complicates analysis of HEIF files, as some data may not be present within a single HEIF file. This poses both complications for forensic previewing/analysis tools, as well as pragmatic concerns for the investigation if external sources cannot be recovered. This particular feature was not examined in-depth during this work, and requires further research, particularly to identify its limitation, such as when specifying networked resources via the URL.

### H. Issues Generating Forensic Hashes

As HEIF is a flexible container format, the same content could be represented in many different ways. While there are limitations on which boxes can be nested, the position of an individual image, or sequence, may be modified, such that images appear in a different order. This should also be possible for tiles comprising a single derived image. In the case of image sequences, there is the option to compress the files using inter-coding, or intra-coding, meaning that the exact same pixel content would have different binary representations. When making use of inter-coding, inserting a new frame will have a substantial impact on the encoding process for all images, resulting in large binary changes for all images in the sequence. Additionally, as the format already supports multiple images, hashing the entire container may have limited use when searching for illegal media, and should be treated akin to hashing a compressed archive, such as a zip file.

One useful feature of HEIF is that it can embed data integrity checks within the `meta` box of a track, inside the DataIntegrityItem of item type `mint`, which can contain an MD5 (`md5i`) checksum of the track content [7]. However, none of the test files in this work contained this item.

### I. Lossless Editing

While HEIF supports lossless editing, in practice this appears to be difficult to achieve at the moment. Experiments showed that when a HEIF file is rotated on an iPhone Max XS, it exports as JPEG afterwards, which also generates an associated .AAE (Apple plist) file containing the modifications. However, when the changes are reverted on the original media, the unmodified HEIF file can again be exported. The authors were unable to find any tool which currently supports lossless editing, instead relying on the Nokia API [16] to facilitate such modifications. At the time of writing, this means that encountering a losslessly edited HEIF file is likely going to be rare. Exported JPEGs which were originally in the HEIF format carry over the metadata, embedding it in the exported HEIF file, which still allows for device source identification. In the future, when this functionality is more easily accessible, the forensic examiner must take care to avoid simply trusting cover and derived images, and should inspect the underlying master images.

## V. RECOMMENDATIONS FOR THE FUTURE

### A. Forensics Methodology

Given that forensic tooling support for HEIF is poor at the moment, it is critical that investigators are provided with a starting point for analysing this format in order to reduce the chance of missing key evidence. As such, brief guidelines are provided here in order to make the best of an unfavourable situation. Ideally, a HEIF specific analysis tool should be created, the requirements of which are discussed in Section V-B, below.

As previously noted, the HEIF format does not have a traditional file signature, meaning that identifying files without an appropriate extension, or verifying the extension, is not straightforward. Instead, the `ftyp` box must be located and the brands matched with appropriate HEIF brands, which will contained the 'heic' brand for all of our test images, though this may not always be the case in reality. The 'pict' type in `hdlr` is also a good indication.

Once it is ascertained that the file is in the HEIF format, the box structure of the file should briefly be analysed, which can either be done using a hex viewer, or by a script which parses the file for strings of length four. Existing ISOBMFF parsers have trouble with some HEIF images, so a simple Python script to extract strings and filter for known ISOBMFF/HEIF boxes works well here. At this stage, the important things are to check the number and type of the `infe` items in the `meta` box at the head of the file, and to determine if there are any `moov` boxes for image sequences.

When analysing `infe` boxes, their role is fairly clear, and displayed as ASCII, as can be seen in Figure 1. An image type such as 'hvc1HEVC Image' indicates a master image, while derived images contain the text 'Derived Image' prefixed by a derived type, such as 'grid' for grids and 'iovl' for alpha and overlays. At this stage the images can be previewed in macOS Preview or GIMP, but care must be taken to verify that the same number of images are present. If not, then the 8th byte of each `infe` item (which includes the 'infe' ASCII), corresponding to the flags component of the box, should be inspected to check if the hidden bit is set. In cases where this bit is set, a copy of the file should be made, and the entry should be changed to visible by entering '00' in place of '01' for all hidden items. For iPhone images, the expectation is that all tiles are hidden and only the primary, derived, image is displayed. However, this can act as a smokescreen to hide an illegal image, such that **it is essential that all hidden images should be toggled to visible** and viewed for verification purposes.

Image sequences, stored in `moov` boxes, are much more complicated as there are many sub-boxes and data structures, and there is also the possibility of a `trak` to be set as inactive. More work needs to be done in this area, however previewing the sequence as a video in a media player, such as MPC, or exporting frames via FFmpeg, both seem to be the most effective methods at the moment. In our tests neither of these seemed to be affected by setting tracks to inactive, however more work is required to explore this more fully, particularly for multi-sequence files, which were not tested here.

In the case of both still images and sequences, the options for previewing metadata and thumbnails appears to be poor. The thumbnail for a cover can be previewed using the paid version of the HEVC extension on Windows 10, or by using the CopyTrans utility on Windows, which both use the embedded thumbnail for the Explorer thumbnail. However, if multiple thumbnails are present in the file, or a thumbnail track is present for sequences, there doesn't appear to be a good way to view these at present without deconstructing the file using one of the C++ APIs. Similarly, while Exiftool is able to preview metadata for the primary item in the file, other metadata elements are difficult to explore, however in this case they can be extracted as text items from the file.

A further complication is the viewing of auxiliary files, which are effectively greyscale images. None of the tested tools made these available, which again would require some programming knowledge in order to deconstruct the file for previewing. Finally, care should be taken to determine if any external file sources have been specified in the `dref` box by manually checking the URL/URN values. This will avoid evidence being missed if it is pulled from a source which is not contained within the HEIF file.

*B. Future Forensics Tooling*

The process above will allow for a reasonable analysis of the data, however there are obvious gaps when analysing image sequences, metadata structures, thumbnails and auxiliary files. Additionally, it is difficult to ascertain the relationship between files and metadata at present without resorting to building scripts to deconstruct the file using an existing API.

In practice, forensic investigators are under too much time pressure to spend an inordinate amount of time analysing a single file. As such, it is clear that a dedicated tool for parsing, previewing, and analysing HEIF files is necessary. A first step would be to deconstruct all elements in the file using existing APIs, which could then be treated as separate files, however this still places a burden on the investigator. The list below provides an ideal set of functionality required of a thorough analysis tool, which should be able to:

- Display all images and image sequences in a HEIF file, regardless of their role and whether they are set to inactive/hidden. In particular this should include thumbnails and auxiliary items.
- Display metadata for all images, tracks and components of a HEIF file. This should also include any relevant flags and metadata included in the HEIF box structure for the items.
- Retrieve and display content pulled in from sources external to the HEIF file via references in the `dref` URL/URN.
- Provide a visual overview of the file structure, such as what images are present, how they are related, encodings used, data offsets, attached metadata and thumbnail items, and any other relevant information.
- Provide hash digests of all images in the file, including both binary and pixel level hashes to account for encoding differences. Individual files in an image sequence should be reconstructed and hashed to avoid attacks using redundant frames inserted into the inter-coded sequence. The use of perceptual hashing may also prove appropriate here.

## VI. CONCLUSIONS

The HEIF format represents a new challenge in the field of digital forensics, requiring new tools and approaches for handling evidence in this format. HEIF is an advanced container format which does not bear much resemblance to existing still image file formats, allowing for multiple images and sequences to be included and arranged in a variety of ways. Critically, HEIF also allows for many embedded items and hidden content, which is difficult to preview as, at the time of writing, support for the format was generally found to be poor. This paper provides a briefing for forensics practitioners on what to expect from HEIF, with a first analysis of its data hiding potential, as well as a first approach to forensic analysis of the format. However, it is critical that future work builds new tooling to facilitate detailed analyses of HEIF file content, as investigators are at risk of missing critical evidence if nothing is done.

TABLE I
HEIF PREVIEWING SUPPORT FOR SELECTED SOFTWARE. WINDOWS 10 RESULTS DIFFER BASED ON THE HEVC EXTENSION VERSION (FREE - 'FROM DEVICE MANUFACTURER', OR PAID VERSION).

| Software | Still Cover Image | | | | | Overlay | Alpha | Additional Still Images | | | | Burst Sequence | | Animated Sequence | |
|---|---|---|---|---|---|---|---|---|---|---|---|---|---|---|---|
| | Display | Crop | Mirror | Rotate | Grid | | | Master | Aux | Thumb | Hidden | Cover | All | Cover | All |
| Win10 HEVC Free | ✓ | ✓ | ✓ | ✓ | ✓ | ✗ | ✗ | ✗ | ✗ | ✗ | ✗ | ✗ | ✗ | ✗ | ✗ |
| Win10 HEVC Paid | ✓ | ✓ | ✓ | ✓ | ✓ | ✓ | ✗ 1st img | ✓ | ✗ | ✗ | ✗ | ✗ | ✗ | ✗ | ✗ |
| CopyTrans HEIC Windows | ✓ | ✗ | ✗ | ✓ | ✓ | ✗ | ✗ | ✗ | ✓ | ✓ | ✗ | ✗ | ✗ | ✗ | ✗ |
| Dropbox Preview | ✓ | ✗ | ✗ | ✓ | ✓ | ✗ | ✗ | ✗ | ✗ | ✓ | ✗ | ✓ | ✓ | ✓ | ✗ |
| macOS Preview 10.14.5 | ✓ | ✓ | ✗ | ✗ | ✓ | ✗ no derived | ✗ no derived | ✓ | ✗ | ✗ | ✗ | ✓ | one per info | ✓ | ✗ |
| macOS Preview 10.15.2 | ✓ | ✓ | ✗ | ✗ | ✓ | ✓ | ✓ | ✓ | ✗ | ✓ | ✗ | ✓ | ✓ | ✓ no animation | ✓ |
| Photoshop Mac 2020 21.0.2 | ✓ | ✓ | ✓ | ✗ | ✗ | ✓ 1st img | ✗ 1st img | ✗ | ✗ | ✗ | ✗ | ✓ | 1st frame | ✓ | ✗ |
| Gimp 2.10.8 Windows | ✓ | ✗ | ✓ | ✗ | ✓ | ✓ | ✓ | ✓ | ✗ | ✗ | ✗ | ✓ | one per info | ✓ | ✗ |
| FFmpeg | ✗ | ✗ | ✗ | ✗ | ✗ | ✗ | ✗ | ✗ | ✗ | ✗ | ✗ | ✓ | ✓ | ✓ | ✓ |

## ACKNOWLEDGEMENT

The authors would like to thank James Thomson (TLA Systems) for his assistance with portions of this work.